\documentclass[prc,aastex,amsmath,amssymb,twocolumn,widetext,floatfix,aps,showpacs]{revtex4-2} 

\usepackage{graphicx}% Include figure files
\usepackage{bm}% bold math
\usepackage{textcomp}
\usepackage{xcolor}
\definecolor{myblue}{rgb}{0.0, 0.0, 0.6}
\usepackage{hyperref}
\hypersetup{
  colorlinks = true,
  citecolor  = myblue,
  linkcolor  = myblue,
  urlcolor   = myblue
}
\usepackage[utf8]{inputenc}
\begin{document}

\title{
  Improved Analysis of the Breakup Corrections to the High Energy $^3$He Beam Polarization measurements with HJET}%

\author{A.~A.~Poblaguev}\email{poblaguev@bnl.gov}

\affiliation{%
 Brookhaven National Laboratory, Upton, New York 11973, USA
}%
\date{May 22, 2023}

\begin{abstract}
 The requirements for hadron polarimetry at the future Electron Ion Collider (EIC) include measurements of the absolute helion ($^3$He, $h$) beam polarization with systematic uncertainties better than $\sigma^\text{syst}_P/P\le1\%$. Recently, it was suggested to utilize the Polarized Atomic Hydrogen Gas Jet Target (HJET) for precision measurement of the polarization of the $\sim$100\,GeV/n helion beam. At the Relativistic Heavy Ion Collider, HJET serves to determine absolute proton beam polarization with low systematic uncertainties of about $\delta^\text{syst}P/P\lesssim0.5\%$. To adapt the HJET method  for the EIC helion beam, the experimentally determined ratio of the beam and target (jet) spin correlated asymmetries should be adjusted by the ratio of $p^\uparrow{h}$ and $h^\uparrow{p}$ analyzing powers $A_\text{N}^{ph}(t)/A_\text{N}^{hp}(t)$ which, in the leading order approximation, is predefined by magnetic moments of the proton and helion, $(\mu_p-1)/(\mu_h/2-1/3)$. However, to achieve the required accuracy in the measured polarization, the corrections due to hadronic spin-flip amplitudes and due to   possible beam $^3$He breakup should be considered. Here a more accurate analysis of the possible breakup corrections to the measured $^3$He beam polarization is provided. The results confirm that the breakup corrections are negligible for the EIC helion beam absolute polarization measurement by HJET.
\end{abstract}

\maketitle

\section{Introduction}

The physics program requirements\,\cite{AbdulKhalek:2021gbh} for hadron polarimetry at the Electron-Ion Collider (EIC)  \cite{Accardi:2012qut} include a precision determination of the $^3$He ($A_h\!=\!3$, $Z_h\!=\!2$) beam polarization,
\begin{equation}
  \sigma_P^\text{syst}/P \lesssim 1\%.
  \label{eq:systEIC}
\end{equation}

It has been advocated\,\cite{Poblaguev:2022gqy,Poblaguev:2022hsi} that the Atomic Polarized Hydrogen Gas Jet Target (HJET)\,\cite{Zelenski:2005mz} can be used for this purpose.

\begin{figure}[b]
  \begin{center}
  \includegraphics[width=0.9\columnwidth]{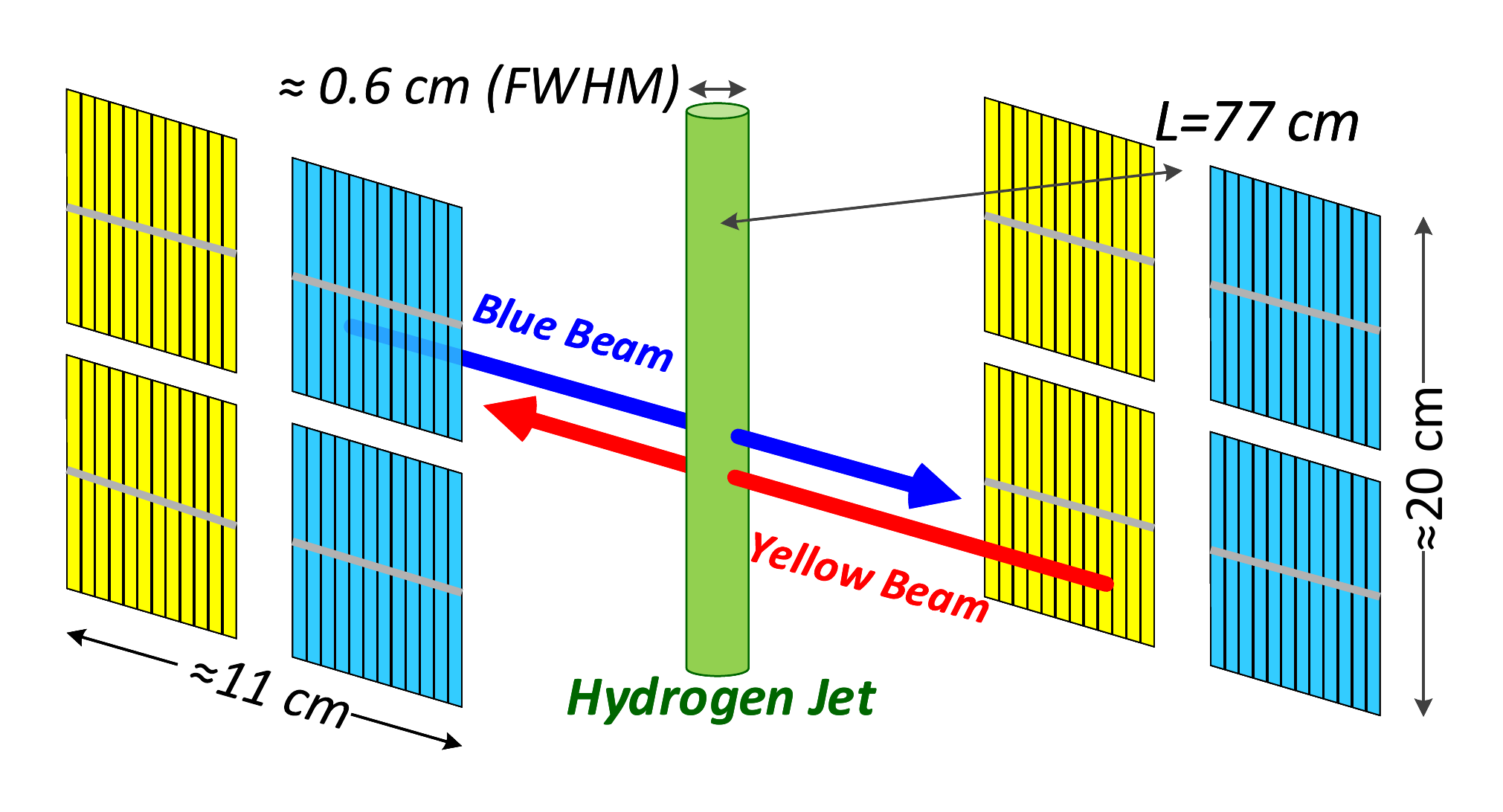}
  \end{center}
  \caption{\label{fig:HJET}
    A schematic view of the HJET recoil spectrometer.
  }
\end{figure}

In the Relativistic Heavy Ion Collider (RHIC) Spin Program\,\cite{Bunce:2000uv}, HJET is employed to measure the absolute transverse (vertical) polarization of the proton beams with a low systematic uncertainty of about $\sigma_P^\text{syst}/P \lesssim 0.5\%$\,\cite{Poblaguev:2020qbw}. Recoil protons from the RHIC beams scattering off the jet target are counted in the left-right symmetric Si strip detectors of the recoil spectrometer shown in Fig.\,\ref{fig:HJET}. For each proton detected, time of flight, kinetic energy $T_R$, and coordinate $z_R$, discriminated by the Si strips, are determined, allowing one to reliably isolate elastic events and subtract the background from the elastic data. The spectrometer geometry predetermines the detection of the recoil protons only in the Coulomb-nuclear interference (CNI) scattering constrained by 
\begin{equation}
  0.0013<-t<0.018\,\rm{GeV}^2.
  \label{eq:tRange}
\end{equation}
The Lorentz invariant momentum transfer $t$ can be simply related to the recoil proton energy,
\begin{equation}
  -t = 2m_pT_R,
\end{equation}
with $m_p$ being the recoil particle (proton) mass.

In HJET measurements, the beam $a_\text{N}^\text{beam}(T_R)$ and target (jet) $a_\text{N}^\text{jet}(T_R)$ spin asymmetries are generally functions of $T_R$. However, since both $a_\text{N}^\text{beam}$ and $a_\text{N}^\text{jet}$ are concurrently calculated using the same elastic events, the found average values of the asymmetries can be used to relate the beam polarization, 
\begin{equation}
  P_\text{beam}=P_\text{jet}\times%
  \big\langle{a_\text{beam}(T_R)}\big\rangle/%
  \big\langle{a_\text{jet}(T_R)}\big\rangle,
  \label{eq:PbeamHJET}
\end{equation}
to the jet one, $P_\text{jet}\approx0.96\pm0.001\%$\,\cite{Zelenski:2005mz}, which is accurately monitored by a conventional  Breit\:\!--\:\!Rabi polarimeter\,\cite{Baumgarten:2001ym}.

The $^3$He${}^\uparrow$ beam polarization can be measured similarly, but, in this case, the right side of Eq.\,(\ref{eq:PbeamHJET}) should be multiplied by the ratio of the $p^\uparrow{h}$ and $h^\uparrow{p}$ analyzing powers, which in leading order approximation is well defined\,\cite{Buttimore:2009zz}:
\begin{equation}
  \frac{A_\text{N}^{ph}(T_R)}{A_\text{N}^{hp}(T_R)} \approx
   \frac{\mu_p-1}{\mu_h/Z_h-m_p/m_h} = -1.283.
\end{equation}
Here, $\mu_h\!=\!-2.128$ and $\mu_p\!=\!2.793$ are the magnetic moments of the helion and proton, respectively.

However, to achieve the beam polarization accuracy satisfying Eq.\,(\ref{eq:systEIC}), corrections due to a priory unknown proton-helion hadronic spin-flip amplitudes and due to possible beam $^3$He breakup, $h\!\to\!pd$ or $h\!\to\!ppn$, should be applied.

It was shown\,\cite{Poblaguev:2022gqy} that $p^\uparrow{h}$ and $h^\uparrow{p}$ hadronic spin-flip amplitudes can be evaluated, with sufficient accuracy, using the results of the $p^\uparrow{p}$ forward elastic analyzing power measurements\,\cite{Poblaguev:2019saw} at HJET.

For a deuteron beam scattering in HJET, the breakup fraction in the elastic data was evaluated\,\cite{Poblaguev:2022hsi}, using HJET measurements with 10, 20, and 31\,GeV/nucleon unpolarized deuteron beams. The result obtained was then extrapolated to the $^3$He beam scattering. Although the breakup corrections were found to be negligible for the $^3$He beam polarization measurements, this conclusion was largely a guesstimate, since an oversimplified and unjustified theoretical model was used to interpret the deuteron data and to make the extrapolation.

Here, a more accurate estimate of the breakup corrections will be provided. The carried-out analysis confirms the conclusion of Ref.\,\cite{Poblaguev:2022hsi} that the EIC $^3$He beam polarization can be measured by HJET with low systematic uncertainties (\ref{eq:systEIC}).

\section{Spin Asymmetries in the Proton-Helion Scattering}

For high energy forward elastic scattering of spin-1/2 particles, a theoretical understanding of the analyzing power structure was developed in Refs.\,\cite{Kopeliovich:1974ee,Buttimore:1978ry,Buttimore:1998rj}. In terms of the spin-flip $\phi_5$ and non-flip $\phi_+$ helicity amplitudes\,\cite{Buttimore:1998rj}, proton-proton CNI analyzing power can be approximated as
\begin{align}
  A_\text{N}(t) &= \frac%
  {-2\rm{Im}\left[
    \phi_5^\text{em} \phi_+^\text{had\,*}\!+%
    \phi_5^\text{had}\phi_+^\text{em\.*}\!+% 
    \phi_5^\text{had}\phi_+^\text{had\,*}\right]}%
  {|\phi_+^\text{had}\!+\!\phi_+^\text{em}|^2}
\nonumber \\
  &= \frac%
  {\sqrt{-t}}{m_p}\,\frac{\kappa_p\,t_c/t -2I_5\,t_c/t - 2R_5}%
  {(t_c/t)^2-2(\rho+\delta_C)\,t_c/t+1},
  \label{eq:AN}
\end{align}
where \textit{had} and \textit{em}\rq\rq\ denote hadronic and electromagnetic parts of the amplitudes, $\kappa_p\!=\!\mu_p\!-\!1\!=\!1.793$, $t_c\!=\!-8\pi\alpha/\sigma_\text{tot}$, $\sigma_\text{tot}$ is the total $pp$ cross section, $\rho$ is the Re/Im amplitude ratio, $\delta_C\!\approx\!\alpha\ln{t_c/t}\!+\!0.024$ is the Coulomb phase\,\cite{Cahn:1982nr,Kopeliovich:2000ez}, and $r_5\!=\!R_5\!+\!iI_5$  is the hadronic single spin-flip amplitude parameter\,\cite{Buttimore:1998rj}, $|r_5|\!\sim\!0.02$\,\cite{Poblaguev:2019saw}. For sake of simplicity, some small, but already essential for the HJET experimental accuracy corrections\,\cite{Poblaguev:2019vho,Poblaguev:2021xkd} were omitted in Eq.\,(\ref{eq:AN}).

For the RHIC proton beam energies, $\rho$  and $\sigma_\text{tot}$ are well known from the world data\,\cite{Workman:2022ynf,COMPETE:2002jcr} fit. For example, for 100\,GeV (lab. system) $\mathit{pp}$ scattering, $\rho\!=\!-0.079$, $\sigma_\text{tot}\!=\!39.4\,\text{mb}$\,\cite{Fagundes:2017iwb}, and $t_c\!=\!-0.0018\,\text{GeV}^2$. Therefore, in HJET measurements, $r_5$ can be derived from the measured $a_\text{N}^\text{jet}(T_R)$. On the other hand, if $r_5$ is already determined elsewhere for the studied beam energy, only $a_\text{N}^\text{beam}(T_R)$ may be measured to calculate the beam polarization.

Since high energy proton-helion experimental data is not yet comprehensive, the measured asymmetry ratio should be used to find the $^3$He beam polarization:
\begin{equation}
  P_h = P_\text{jet}\frac{a_\text{beam}(T_R)}{a_\text{jet}(T_R)}\times%
  \frac%
  {\kappa_p - 2I_5^{ph} - 2R_5^{ph}\,T_R/T_c}%
  {\kappa_h - 2I_5^{hp} - 2R_5^{hp}\,T_R/T_c},
  \label{eq:Ph}
\end{equation}
where $\kappa_h\!=\!\mu_h/Z_h\!-\!m_p/m_h\!=\!-1.398$, $r_5^{ph}$ and $r_5^{hp}$ are hadronic spin-flip amplitude parameters for $p^\uparrow h$ and $h^\uparrow p$ scattering, respectively,  and $T_c\!=\!4\pi\alpha{Z_h}/m_p\sigma_\text{tot}^{ph}\!\approx\!0.7\,\text{MeV}$.

Anticipating systematic errors due to possible inaccuracy in the evaluation of $r_5^{ph}$ and $r_5^{hp}$ and the breakup corrections, the measured beam polarization,  as a function of $T_R$, can be related to the actual $^3$He polarization, $P_h$, as
\begin{align}
  P_\text{meas}(T_R) &= P_h\times\left[1+\xi(T_R)\right],
  \label{eq:Pmeas}\\
  \xi(T_R) &\approx \xi_0 + \xi_1\,T_R/T_c.
  \label{eq:xi}
\end{align}

Extrapolating $P_\text{meas}(T_R)$ to $T_R\!\to\!0$, one can determine the beam polarization with a systematic uncertainty given only by errors in the evaluation of $I_5^{ph}$ and $I_5^{hp}$
\begin{equation}
  P_h=P_\text{meas}(0),\quad%
    \delta^{r_5} P_h/P_h = \frac{2}{\kappa_p}\delta I_5^{ph}%
    - \frac{2}{\kappa_p}\delta I_5^{hp}.
\end{equation}

\section{Hadronic Spin-Flip Amplitudes in High Energy Proton-Nucleus Scattering}

  In Ref.\,\cite{Kopeliovich:2000kz}, it was shown that, at high energy, to a very good approximation, the ratio of the spin-flip to the non-flip parts of the elastic proton-nucleus amplitude is the same as for proton-nucleon scattering. In terms of the hadronic spin-flip amplitude parameter $r_5$, the result can be written as
\begin{equation}
  r_5^{pA} = \frac{i+\rho^{pA}}{i+\rho}r_5\approx r_5,
  \label{eq:r5_pA}
\end{equation}
where  $\rho^{pA}$ is the real-to-imaginary ratio for the elastic $pA$ scattering. It may be noted that Eq.\,(\ref{eq:r5_pA}) can be readily derived considering the polarized proton scattering off an unpolarized nucleus in Glauber (diffraction) approximation\,\cite{Glauber:1955qq,*Glauber:1959}.

In this approach, elastic ($f\!=\!i$) and/or breakup ($f\!\ne\!i$) proton-nucleus amplitude can be presented\,\cite{Glauber:1970jm} using the following integral over the impact vector $\bm{b}$:
\begin{align}
  F_{fi}(\bm{q}) &= \frac{ik}{2\pi}\int{d^2\bm{b}\,e^{i\bm{qb}}}\,%
    \prod_{j=1}^A{d^3\bm{r}_j}% 
    \nonumber \\ &\qquad\qquad%
    \Psi_f^*(\{r_j\})\Gamma(\bm{b},\bm{s}_1...\bm{s}_A)\Psi_i(\{r_j\})%
    \label{eq:Ffi}
\end{align}
where $\bm{k}$ is the momentum of the incident proton, $\Psi_i$ and $\Psi_f$ are the nucleus's initial and final state wave functions, and $\Gamma(\bm{b},\bm{s}_1...\bm{s}_A)$ is the profile function. The positions of the $A$ nucleons in the nucleus were defined by the vectors $\bm{r}_j,~j\!=\!1,\dots,A$, and $\bm{s}_j$ are the projections of these vectors on the plane perpendicular to $\bm{k}$.

For a proton-deuteron small angle scattering, the elastic $pd$ amplitude $F_{ii}$ can be approximated\,\cite{Franco:1965wi} via proton-proton $f_p$ and proton-neutron $f_n$ ones as 
\begin{align}
  F_{ii}(\bm{q}) &=
  S(\bm{q}/2)f_n(\bm{q}) +  
  S(\bm{q}/2)f_p(\bm{q})
  \nonumber \\ &+
  \frac{i}{2\pi k}\int{S(\bm{q}')f_n(\bm{q}/2\!+\!\bm{q}')f_p(\bm{q}/2\!-\!\bm{q}')d^2\bf{q}'},
  \label{eq:Fii_pd}
\end{align}
where $S(\bm{q})$ can be interpreted as a deuteron form factor.

To calculate the $p^\uparrow d$ spin-flip amplitude $F_{ii}^\text{sf}$, one can utilize the proton-nucleon one, $f_N^\text{sf}(\bm{q})$, which, according to the definition of $r_5$\,\cite{Buttimore:1998rj}, is
\begin{equation}
  f_N^\text{sf}(\bm{q})=\frac{\bm{qn}}{m_p}\,\frac{r_5}{i+\rho}\,f_N(\bm{q})
  = \bm{qn}\,\hat{r}_5f(\bm{q}).
\end{equation}
Since in the HJET measurements $|r_5|\,q/m_p\lesssim0.003$, one and only one non-flip amplitude $f_N$ in each term of sum (\ref{eq:Fii_pd}) should be replaced by $f_N^\text{sf}$. In particular,
\begin{align}
  f_pf_n &\to \left[(\bm{q}/2\!+\!\bm{q}')\bm{n}f_p\,f_n + f_p\,(\bm{q}/2\!-\!\bm{q}')\bm{n}f_n\right]\,\hat{r}_5
 \nonumber \\ &=
 \bm{qn}\,\hat{r}_5\,f_pf_n.
\end{align}
Since each term on the right side of Eq.\,(\ref{eq:Fii_pd}) acquired a factor $\bm{qn}\,\hat{r}_5$, one immediately arrives at Eq.\,(\ref{eq:r5_pA}).

Generally, elastic $pA$ amplitude can be expressed as 
\begin{align}
  F_{ii}(\bm{q}) &=
  \sum_i{\{\mathcal{S}_if_i\}} + 
  \sum_{i,j}{\{\mathcal{S}_{ij}f_if_j\}}
  \nonumber \\ &+ 
  \sum_{i,j,k}{\{\mathcal{S}_{ijk}f_if_jf_k\}} + \dots,
  \label{eq:Fii_pA}
\end{align}
where, e.g., for the 3-amplitude terms,
\begin{align}
  \{\mathcal{S}_{ijk}f_if_jf_k\} &=
  \int{\mathcal{S}_{ijk}(\bm{q}_i',\bm{q}_j',\bm{q}_k')\,%
    f_i(\bm{q}_i')f_j(\bm{q}_j')f_j(\bm{q}_k')}%
    \nonumber \\ &\times%
    \delta(\bm{q}-\bm{q}_i'-\bm{q}_j'-\bm{q}_k')d^2\bm{q}_i'd^2\bm{q}_j'd^2\bm{q}_k'  
\end{align}
and similarly for other terms. Thus, no detailed knowledge of the form factor functions $\mathcal{S}_{i{\dots}j}$ is needed to prove Eq.\,(\ref{eq:r5_pA}).

\begin{figure}[t]
  \begin{center}
  \includegraphics[width=0.4\columnwidth]{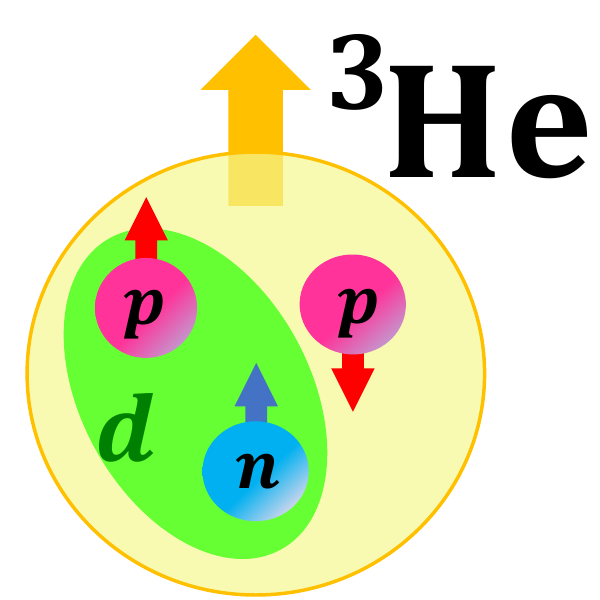}
  \end{center}
  \caption{\label{fig:He3}
    A polarized $^3$He spin structure in the ground ${}^1S_0$ state. Due to the Pauli principle, the protons are in a spin singlet state. A bound state of the neutron and proton with parallel spins may be approximated by a deuteron. 
  }
\end{figure}

In the case of unpolarized proton scattering of a fully polarized nucleus with all nucleons having the same polarization $P$, one easily finds $r_5^{Ap}\!\cong\!r_5P$. In another obvious case of a space symmetric distribution of the nucleons, the result is proportional to the average polarization of the nucleons, $r_5^{pA}\!\cong\!r_5\sum{P_i}/A$. Assuming that $^3$He nuclei are in the space-symmetric ${}^1S_0$ state (Fig.\,\ref{fig:He3}), in which the helion spin is carried by the neutron (i.e., $P_n\!=\!1$ and $P_p\!=\!0$), one finds $r_5^{hp}\!=\!r_5/A_h$\,\cite{Buttimore:2001df}. Small corrections to $P_{n,p}$ due to the ${}^3S_1$ and ${}^3D_1$ partial waves were evaluated in Ref.\,\cite{Friar:1990vx}.

Thus, the proton-helion hadronic spin-flip amplitudes can be related\,\cite{Poblaguev:2022gqy} to the proton-proton ones,
\begin{equation}
  r_5^{ph}= r_5\frac{i+\rho^{ph}}{i+\rho},\qquad%
  r_5^{hp}\approx0.27r_5\frac{i+\rho^{ph}}{i+\rho},
  \label{eq:r5el}
\end{equation}
with expected calculation accuracy of about $|\delta r_5^{ph,hp}|\lesssim0.1\%$.

Considering a breakup scattering, e.g., $hp\!\to\!dp\:\!p$, one can define the non-flip and spin-flip amplitudes as
\begin{align}
  \!\!F_{fi}(\bm{q}) &= \psi_{fi}(\bm{q})F_{ii}(\bm{q}),\quad
  \psi_{fi}(\bm{q})=|\psi_{fi}(\bm{q})|e^{i\varphi_{fi}(\bm{q})},
  \label{eq:F_fi}
  \\
  \!\!F_{fi}^\text{sf}(\bm{q}) &= \frac{\bm{qn}}{m_p}\,\frac{\widetilde{r}_5}{i+\rho^{pA}}F_{fi}(\bm{q}),
  \label{eq:F_fi_sf}
\end{align}
Here, $F_{fi}(\bm{q})$ and $F_{fi}^\text{sf}(\bm{q})$ are effective breakup amplitudes (for a given value of $\bm{q}$), i.e., sums of all amplitudes over all internal states of the breakup. Therefore, the calculation of probabilities, such as $|F_{fi}|^2$, $|F_{fi}^\text{sf}|^2$, and $\mathrm{Im}\:\!F_{fi}^\text{sf}F_{fi}^*$ assumes summation over these states.

Since, the only difference between elastic and breakup scattering is given [see Eq.\,(\ref{eq:Ffi})] by the final state wave function $\Psi_f$, the amplitude expansion (\ref{eq:Fii_pA}) should be also valid (with some other set of functions $\mathcal{S}_{i{\dots}j}$) for the breakup scattering. Thus,
\begin{equation}
  \widetilde{r}_5^{ph}=r_5\frac{i+\rho^{ph}}{i+\rho},\qquad%
  \widetilde{r}_5^{hp}=(0.27+\delta_{pd})\,r_5\frac{i+\rho^{ph}}{i+\rho}.
  \label{eq:r5brk}
\end{equation}

Considering a single nucleon scattering of an unpolarized proton from the fully polarized helion $h^\uparrow$ in the ground state (see Fig.\,\ref{fig:He3}) and assuming that $\bm{q}$ is sufficiently large to knock out the target nucleon, one should conclude that the $h\to pd$ helion breakup can occur only if the beam proton was scattered off the oppositely polarized proton, $p^\downarrow$, in $^3$He. Such a speculation suggests that $\widetilde{r}_5^{hp}=-r_5$ (or $0.27\!+\!\delta_{pd}\!=\!-1$). As the accordance between the $h\to pd$ breakup and the beam scattering off the $p^\downarrow$ nucleon should be diluted at low $\bm{q}$, it will be assumed
\begin{equation}
  -1.27 \le \delta_{pd} \le 0
\end{equation}
in the estimates below.

\section{Inelastic Scattering in HJET}

Although the scattered beam particles are not observed at HJET, the polarimeter has the capability for the identification of inelastic events\,\cite{Poblaguev:2020qbw}. For a scattered beam particle with mass $M$, the recoil proton coordinate $z_R$ in the detector has the following dependence on the kinetic energy $T_R$ and the effective scattered (missing) mass $M_X$:
\begin{align}
  \!\!\!\frac{z_R-z_\text{jet}}{L}&=%
  \sqrt{\frac{T_R}{2m_p}}\times
  \left[ 1+\frac{m_p^2}{ME_\text{beam}}%
    +\frac{m_p\Delta}{T_RE_\text{beam}}%
    \right]\!\!,
\end{align}
where $E_\text{beam}$ is the beam energy\,per\,nucleon, $\Delta\!=\!M_X\!-\!M$, $z_\text{jet}$ ($\big\langle{z_\text{jet}}\big\rangle\!=\!0$, $\langle{z_\text{jet}^2}\rangle^{1/2}\!\approx\!2.5\,\text{mm}$) is the coordinate of the scattering point, and $L\!=\!77\,\text{cm}$ is the distance to the detectors.

For elastic scattering, $z_R$ linearly depends on $T_R^{1/2}$. The possibility to isolate inelastic events in HJET depends on the value of the $\Delta/E_\text{beam}$ ratio\,\cite{Poblaguev:2022hsi}. If, for a given $T_R$, the corresponding correction to $z_R$ exceeds the smearing of the measured $z_R$ due to the jet thickness, the breakup and elastic events can be separated. However, if $\Delta$ is large enough, the breakup event cannot be detected in HJET.

As it was illustrated in Ref.\,\cite{Poblaguev:2022xoa}, inelastic $pp$ scattering ($\Delta\ge m_\pi$) is almost invisible in HJET for $E_\text{beam}\!=\!100\,\text{GeV}$, but may be detected and well separated from the elastic events if $E_\text{beam}\!=\!255\,\text{GeV}$.

In numerous gold beam (3.9\:\!--\:\!31\,GeV/nucleon energy range) measurements at HJET\,\cite{Poblaguev:2022xoa}, pion production events could not be detected, but a high rate of the breakup events ($\Delta_\text{Au}\gtrsim4\,\text{MeV}$) was anticipated. However, no breakup events were reliably identified in these data\,\cite{Poblaguev:2022hsi}.

The result can be readily explained, if the breakup is going via incoherent scattering of the beam (jet) proton off a nucleon in the nucleus. In this case\,\cite{Poblaguev:2022hsi},
\begin{equation}
  \Delta = \left(1-\frac{m_p}{m_A}\right)T_R+p_x\sqrt{\frac{2T_R}{m_p}},
  \label{eq:Delta}
\end{equation}
where $p_x$ is the internal motion momentum of the nucleon (in the direction of the detector). Since $T_R\lesssim10\,\text{MeV}$ for recoil protons detected at HJET, the scattered mass excess is much smaller than the nucleus mass, $\Delta\ll M$. Consequently, the breakup event rate should be strongly suppressed by the phase space factor.

\subsection{A Model Used to Isolate Breakup Events in HJET}

To experimentally evaluate the breakup fraction in the elastic data in the deuteron beam measurements at HJET, the following model for the breakup event distribution was suggested in Ref.\,\cite{Poblaguev:2022hsi}.

Assuming that the $p_x$ distribution in a deuteron is given by a unity integral normalized Breit\:\!--\:\!Wigner function%\,\cite{Poblaguev:2022hsi}
\begin{equation}
  f_\text{BW}(p_x,\sigma_{p}) = \frac{\pi^{-1}\,\sqrt{2}\sigma_{p}}{p_x^2+2\sigma_{p}^2},
  \label{eq:fBW}
\end{equation}
the $d\to pn$ breakup fraction $\omega(T_R,\Delta)$ can be found\,\cite{Poblaguev:2022hsi} as a convolution of the $\Delta$ distribution, calculated in accordance with Eq.\,(\ref{eq:Delta}), and the phase space integral (calculated as a function of $\Delta$),  
\begin{align}
  &\omega(T_R,\Delta) =%
  \frac{d^2\sigma_\text{brk}(T_R,\Delta)}{d\sigma_\text{el}(T_R)\,d\Delta} =%
  \frac{\sqrt{2m_pm_n}}{4\pi m_d}
  \nonumber \\ &\qquad\times%
  |\bar{\psi}(T_R,\Delta)|^2f_\text{BW}(\Delta\!-\!\Delta_0,\sigma_\Delta)%
  \sqrt{ \frac{\Delta\!-\!\Delta^d_\text{thr}}{m_d} },
  \label{eq:omegaTR-D} 
  \\  &
  \Delta_0=(1-m_p/m_d)T_R,\qquad%
  \sigma^2_\Delta=2\sigma_{p}^2T_R/m_p.
  \label{eq:omegaD0}
\end{align}
Since the key dependence of the breakup amplitude on $T_R$ and $\Delta$ is allocated in the function  $f_\text{BW}$, $\bar{\psi}(T_R,\Delta)$ should be interpreted as a reduced ratio of the breakup to elastic amplitudes. Therefore, in HJET measurements, i.e., for low $T_R$ and $\Delta$, $\bar{\psi}(T_R,\Delta)$ can be substituted by a constant $\bar{\psi}\!\equiv\!\bar{\psi}(0,0)$. For the $d\!\to\!pn$ breakup, the threshold is $\Delta^d_\text{thr}\!=\!m_p\!+\!m_n\!-\!m_d\!=\!2.2\,\text{MeV}$.

Introducing
\begin{equation}
  \omega_{\Phi}(T_R) =  \int_{\Delta^d_\text{thr}}^{\infty} {\omega(T_R,\Delta)_{|\bar{\psi}|=1}\,d\Delta},
\end{equation}
which can be calculated within the model used, one can express the breakup fraction dependence on $T_R$ as
\begin{equation}
  \omega(T_R) =%
  \frac{d\sigma_\text{brk}(T_R)}{d\sigma_\text{el}(T_R)} = |\bar{\psi}|^2\omega_\Phi(T_R).
\end{equation}

To search for breakup events in the HJET data, the recoil proton displacement, 
\begin{equation}
  \zeta = z_R^\text{brk}(T_R,\Delta)-\big\langle{z_R^\text{el}(T_R)}\big\rangle%
  \approx \sqrt{\frac{T_R}{2m_p}}\,\frac{L\Delta}{E_\text{beam}},
  \label{eq:zeta}
\end{equation}
relative to the mean elastic coordinate for the same $T_R$ can be considered. Using Eqs.\,(\ref{eq:Delta}) and (\ref{eq:omegaTR-D}), one readily finds the breakup fraction as a function of $\zeta$,
\begin{align}
  \omega_z(T_R,\zeta) &\propto%
  f_\text{BW}(\zeta-\zeta_0,\sigma_z)\sqrt{\zeta-\zeta_\text{thr}^d},
  \\
  \sigma_z &=  L\sigma_p/E_\text{beam},
  \label{eq:sigma_z} \\
  \zeta_0 &=%
  \left(1-\frac{m_p}{m_d}\right)\frac{\sqrt{T_Rm_p/2}}{\sigma_p}\sigma_z%
  \approx 0.62\sigma_z,
  \label{eq:zeta_0} \\
  \zeta_\text{thr}^d &=%
  \sqrt{\frac{m_p}{2T_R}}\frac{\Delta_\text{thr}^d}{\sigma_p}\sigma_z%
  \approx 0.69\sigma_z.
  \label{eq:zeta_thr} 
\end{align}
Numerical estimates in Eqs.\,(\ref{eq:zeta_0}) and (\ref{eq:zeta_thr}) have been done for $T_R\!=\!4\,\text{MeV}$ and $\sigma_p\!=\!35\,\text{MeV}$. Since possible variations of $T_R$ at HJET may lead only to about 50\% alteration of small values of $\zeta_0$ and $\zeta_\text{thr}^d$, for an immediate estimate, the breakup rate dependence on $\zeta$ may be considered as $T_R$-independent.

Obviously, $\omega_z(T_R,\zeta)d\zeta$ distribution is linearly scaled by the beam energy $E_\text{beam}$, e.g., $\sigma_z\!=\!2.7\,\text{mm}$ (10 GeV) and $\sigma_z\!=\!0.27\,\text{mm}$ (100 GeV). However, since the measured $z$-coordinate, elastic or inelastic, is smeared, $\sim$\;\!$\exp{(-z_\text{jet}^2/2\sigma_\text{jet}^2)}$, due to the jet width, the breakup fraction isolation is essentially the beam energy dependent. Also, it should be noted that measured $z_R$ cannot be outside the detector, i.e., it must be (approximately)
\begin{equation}
  \zeta <\zeta^\text{max}\approx 54-18\sqrt{T_R/\text{MeV}}\,\text{mm}.
\end{equation}

\begin{figure}[t]
  \begin{center}
  \includegraphics[width=0.9\columnwidth]{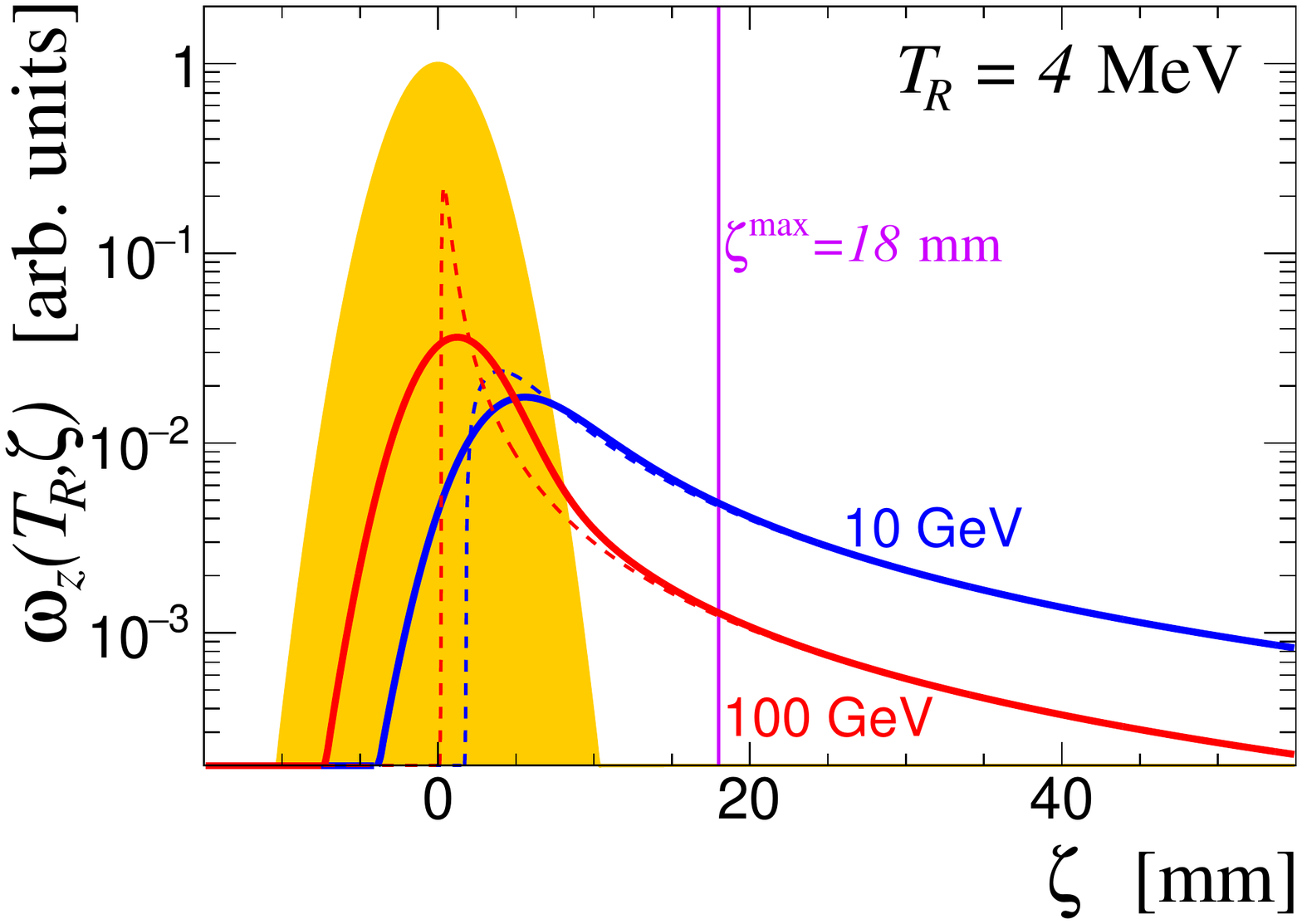}
  \includegraphics[width=0.9\columnwidth]{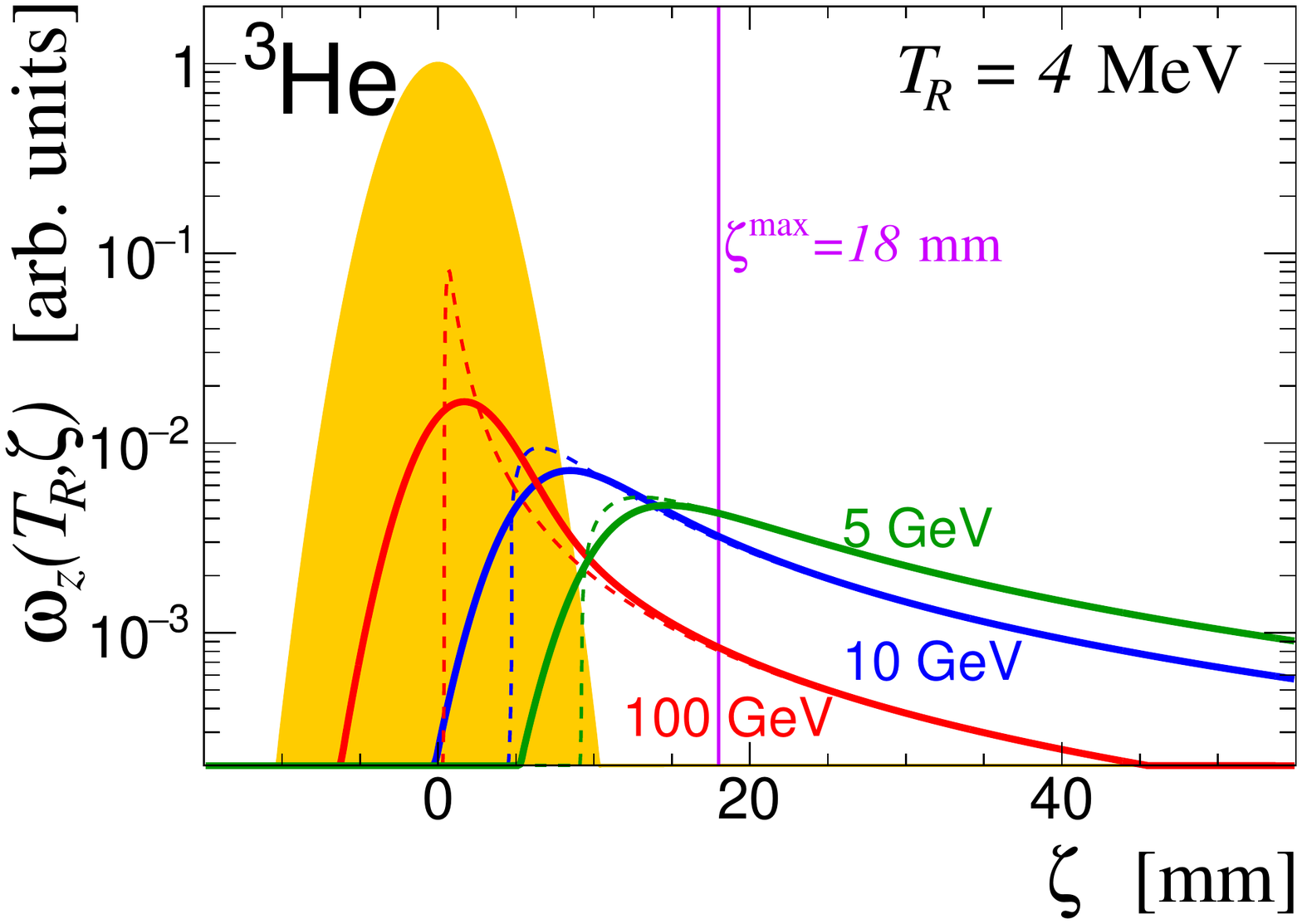}
  \end{center}
  \caption{\label{fig:zeta}
    The model-based calculation of $\omega_z(T_R\!=\!4\,\text{MeV},\zeta)$ for the deuteron and helion beams. Elastic events are shown by filled areas. Solid lines are used to display the breakup distributions for 5 (green), 10 (blue), and 100 (red) GeV/nucleon beam energies. The breakup integrals are 5\% ($^2$H) and 2.5\% ($^3$He) ff the elastic ones. Dashed lines stand for the breakup distributions without $z_\text{jet}$ related smearing.
  }
\end{figure}

A possibility to separate the beam $d\to pn$ breakup events from the elastic ones at HJET is illustrated in Fig.\,\ref{fig:zeta} for two deuteron beam energies, 10 and 100\,GeV/nucleon. One can see, that for $T_R\!=\!4\,\text{MeV}$, the breakup events can be isolated, for both beam energies, if $10\!<\!\zeta\!<\!18\,\text{mm}$. However, in the case of $E_\text{beam}\!=100\,\text{GeV/nucleon}$, (i) there will be a larger rate from the pion production scattering\,\cite{Poblaguev:2022xoa} and (ii) the experimental uncertainties may exceed the breakup component.

\subsection{Evaluation of the $d\to pn$ Breakup Fraction in the Deuteron Beam Measurements at HJET}

\begin{figure}[t]
  \begin{center}
  \includegraphics[width=0.9\columnwidth]{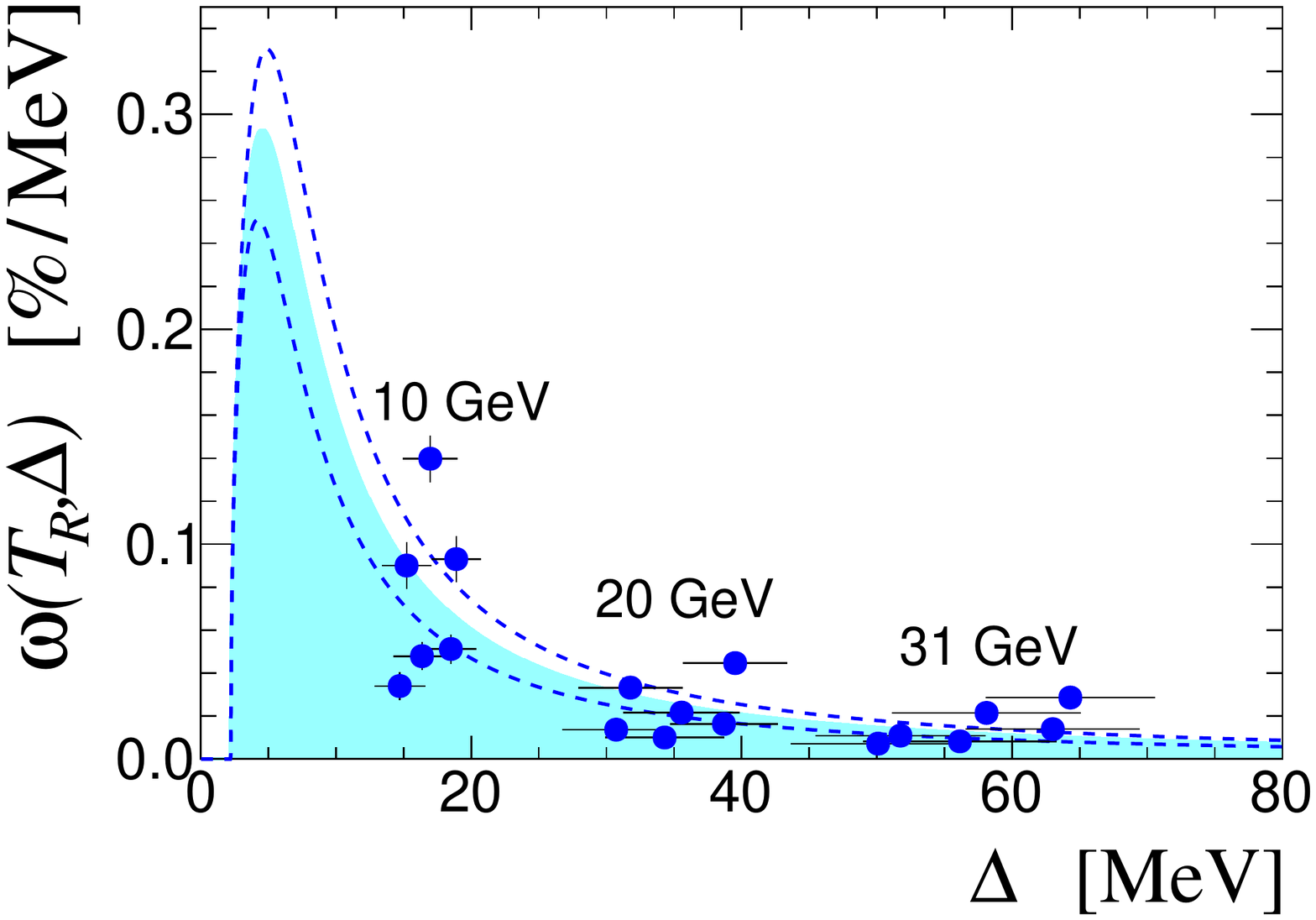}
  \end{center}
  \caption{\label{fig:Run16}
    Experimental evaluation of the $pd\!\to\!ppn$ breakup rate\,\cite{Poblaguev:2022hsi}. The filled distribution ($T_R\!=\!3.5\,\text{MeV}$ and $\sigma_p\!=\!35\,\text{MeV}$) integral is 5\%. Dashed lines are for similarly normalized ($\sigma_p$, $|\bar{\psi}|$) functions (\ref{eq:omegaTR-D}) calculated for 2.8 and 4.2\,MeV kinetic energies.
  }
\end{figure}

In RHIC Run\,16\,\cite{Liu:IPAC2017-TUPVA046}, deuteron\:\!--\:\!gold collisions were studied at several beam energies. Since HJET operated in this unpolarized ion Run, the deuteron beam breakup fraction was experimentally evaluated for three beam energies, 9.8, 19.6, and 31.3\,GeV/nucleon, i.e., integrally in $12\!<\Delta\!<\!72\,\text{MeV}$ scattered mass excess range. The measurements were done in the recoil proton kinetic energy range $2.8\!<\!T_R\!<\!4.2\,\text{MeV}$.

In the data analysis, the breakup fraction was determined\,\cite{Poblaguev:2022hsi} in 18 data bins, discriminated by $T_R$, $\zeta$, and $E_\text{beam}$. Using Eq.\,(\ref{eq:zeta}), $\Delta$ was calculated for each bin and the model parameters
\begin{equation}
  \sigma_p\approx35\,\text{MeV},\qquad  |\bar{\psi}|\approx5.7,
  \label{eq:omegaCalib}
\end{equation}
for the deuteron beam were found in the fit. 

Results of the measurements are illustrated in Fig.\,\ref{fig:Run16}. Obvious difficulty in the interpretation of the fit is that the unverified model (\ref{eq:omegaTR-D}) was used to extrapolate $\omega(T_R,\Delta)$ from large $\Delta$, where it was measured, to low $\Delta$, where the main contribution to $\omega(T_R)$ comes from.

Nonetheless, the value of $\sigma_p$ found in the fit leads, within the model used, to the following slope of the diffraction cone in the elastic $\mathit{pd}$ scattering
\begin{equation}
  B^{pd} = (1-m_p/m_d)^2/8\sigma_p^2 + B = 37\,\text{GeV}^{-2}
\end{equation}
where $B\!=\!11\,\text{GeV}^{-2}$\,\cite{Bartenev:1973jz,*Bartenev:1973kk} is the elastic $\mathit{pp}$ slope. The slope calculated is in reasonable agreement with the experimental values for $B^{pd}$\,\cite{Beznogikh:1973uka,Akimov:1975rm}.

\begin{figure}[t]
  \begin{center}
  \includegraphics[width=0.9\columnwidth]{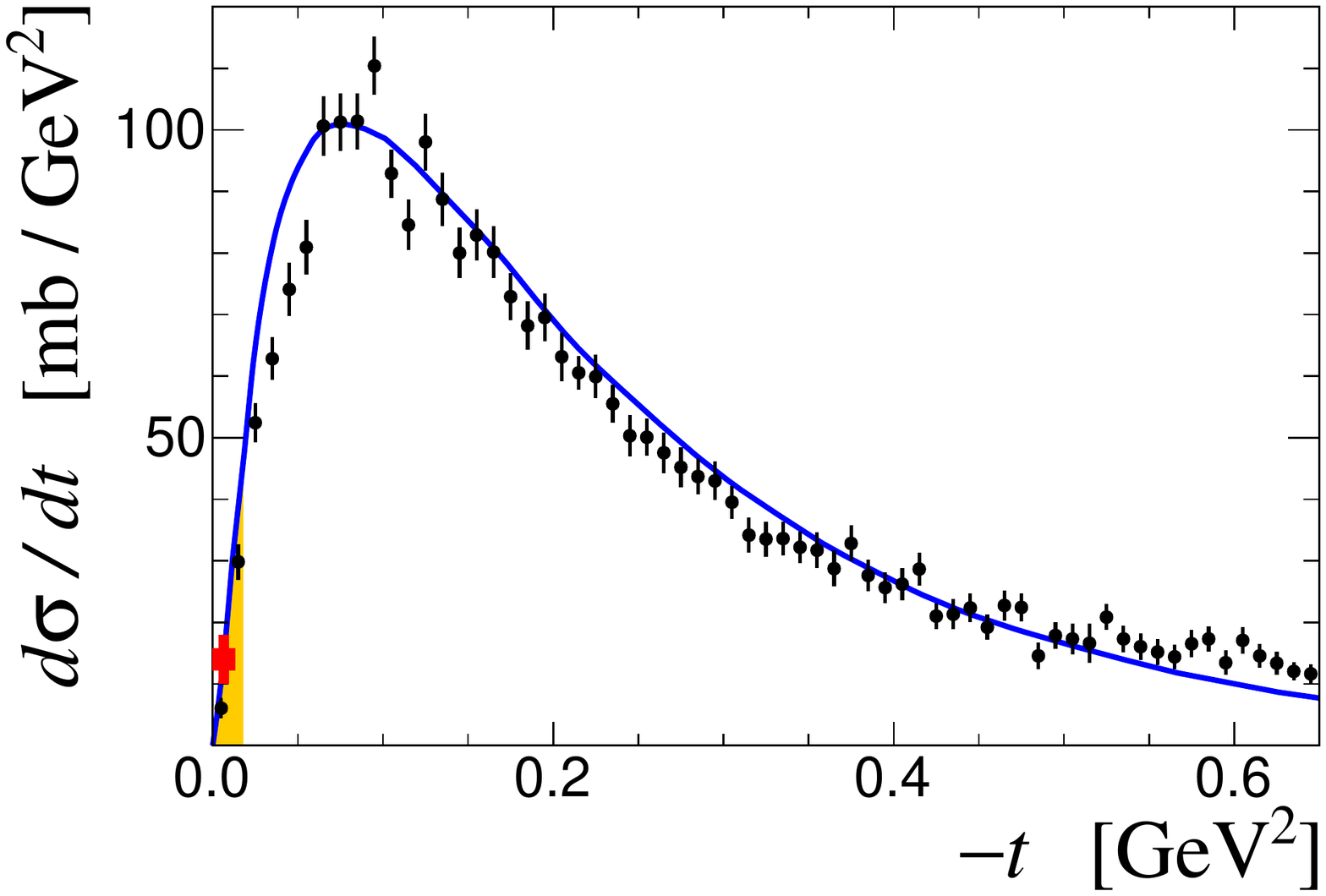}
  \end{center}
  \caption{\label{fig:pd-ppn}
  Differential cross section for the $pd\!\to\!ppn$ breakup scattering. The experimental data ($\bullet$) and theoretical calculation (solid line) are taken from Ref.\,\cite{Aladashvili:1977xe}. The result evaluated in the HJET measurement\,\cite{Poblaguev:2022hsi} is marked {\color{red}$\blacksquare$}. The filled area indicates part of the distribution available for measurement at HJET.
  }
\end{figure}

In the HJET measurements, the breakup fraction in the elastic $\mathit{pd}$ data was estimated\,\cite{Poblaguev:2022hsi} as 
\begin{equation}
  \big\langle\omega_d(T_R)\big\rangle_\text{2.8\:\!--\:\!4.2\,MeV} = 5.0\pm1.4\%.
  \label{eq:dBreakup}
\end{equation}
The value found can be compared with the $pd$\:\!$\to$\:\!$ppn$ differential cross-section (Fig.\,\ref{fig:pd-ppn}) measured in  1.8\,GeV/nucleon deuteron beam scattering in a hydrogen bubble chamber\,\cite{Aladashvili:1977xe}.  Using, for normalization,  elastic $pd$ differential cross section\,\cite{Beznogikh:1973uka}, Eq.\,(\ref{eq:dBreakup}) can be re-written as
\begin{equation}
  d\sigma/dt\big|_{-t=0.0066\,\text{GeV}^2} = 14\pm4\,\text{mb/GeV}^2.
\end{equation}
The result is in good agreement with the final state interaction model calculation\,\cite{Aladashvili:1977xe}, $\approx$\:\!$15\,\text{mb/GeV}^2$, and in fair consistency with the value, $8\pm2\,\text{mb/GeV}^2$, interpolated from experimental points in Fig.\,\ref{fig:pd-ppn}. It should also be pointed out that only a small fraction, $\approx$\:\!$1.5\%$, of all  $pd\to ppn$ events can be detected in HJET. 

\subsection{The Helion Beam Breakup in HJET}

Applying the breakup model, after replacing $m_d\!\to\!m_h$, $m_n\!\to\!m_d$, and $\Delta_\text{thr}^d\!\to\!\Delta_\text{thr}^h\!=\!5.5\,\text{MeV}$ in Eqs.\,(\ref{eq:omegaTR-D}) and (\ref{eq:omegaD0}), for the $^3$He beam and using the parametrization (\ref{eq:omegaCalib}), one can evaluate\,\cite{Poblaguev:2022hsi} the breakup fraction for the HJET momentum transfer range (but disregarding the detector acceptance)
\begin{equation}
  \big\langle\omega_h(T_R)\big\rangle_\text{1\:\!--\:\!10\,MeV} = 2.4\pm0.4\,\%.
  \label{eq:hBreakup}
\end{equation}
Assuming that, for low $t$, the three-body breakup $h\!\to\!ppn$ is much more strongly suppressed by a phase space factor, only the two-body one, $h\!\to\!pd$, was considered.

\begin{figure}[t]
  \begin{center}
  \includegraphics[width=0.9\columnwidth]{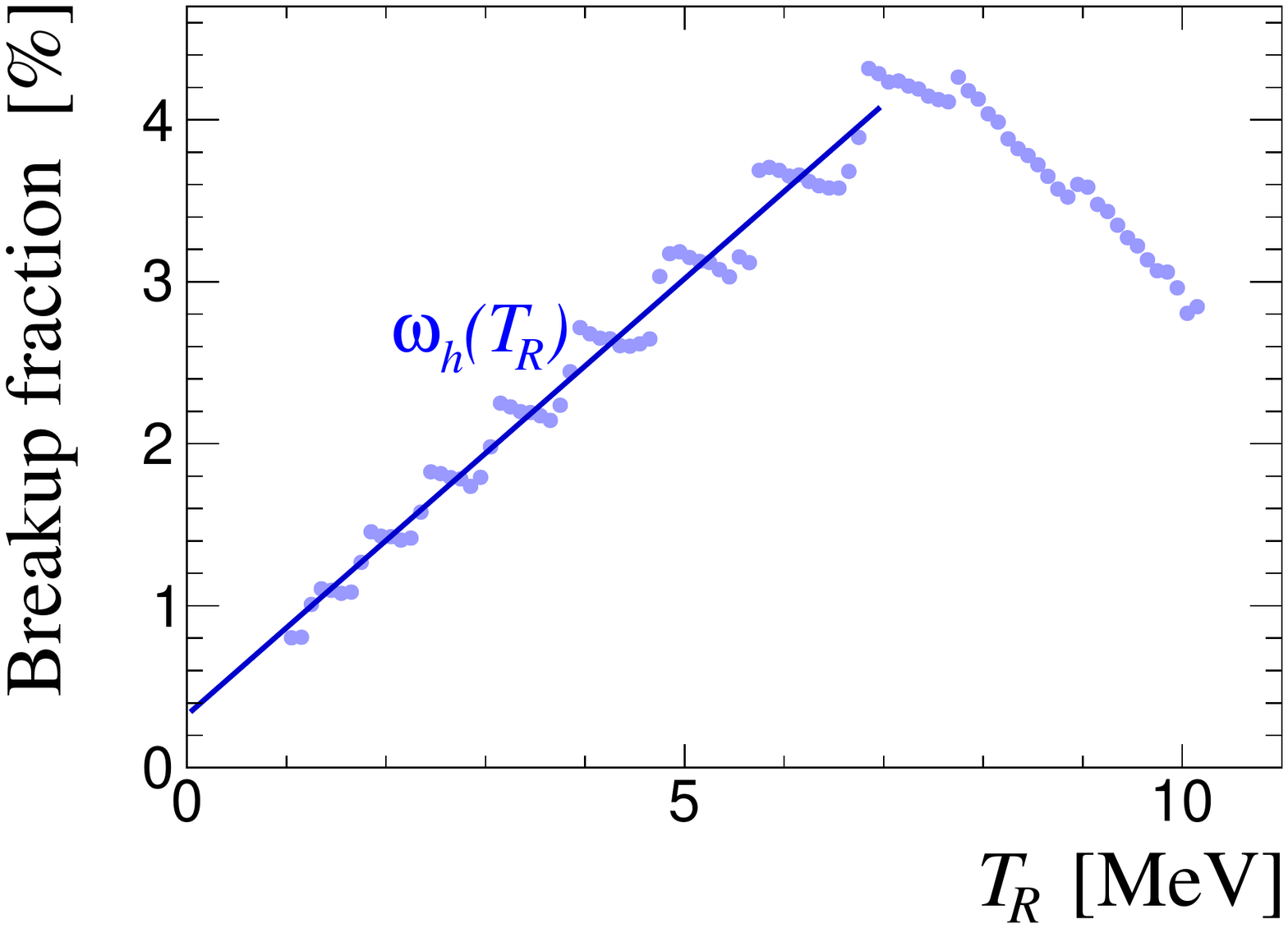}
  \end{center}
  \caption{\label{fig:omega}
    An estimate of the breakup fraction $\omega_h(T_R)$ for 100\,GeV/nucleon $^3$He beam polarization  measurement with HJET. For calculations, the function $f_\text{BW}$ parametrization (\ref{eq:omegaCalib}) found in the deuteron beam data analysis\,\cite{Poblaguev:2022hsi} was used.
  }
\end{figure}

Shown in Fig.\,\ref{fig:omega} is a calculation of the {\emph effective} breakup function $\omega_h(T_R)$ for a 100\,GeV/nucleon helion beam. The non-smooth dependence of the calculated points on $T_R$ reflects the discrete changes in the event selection efficiency attributed to the Si strip width (3.75\,mm). Near 7\,MeV, the linear dependence on $T_R$ is broken due to the finite size of the detector.

\begin{figure}[t]
  \begin{center}
  \includegraphics[width=0.9\columnwidth]{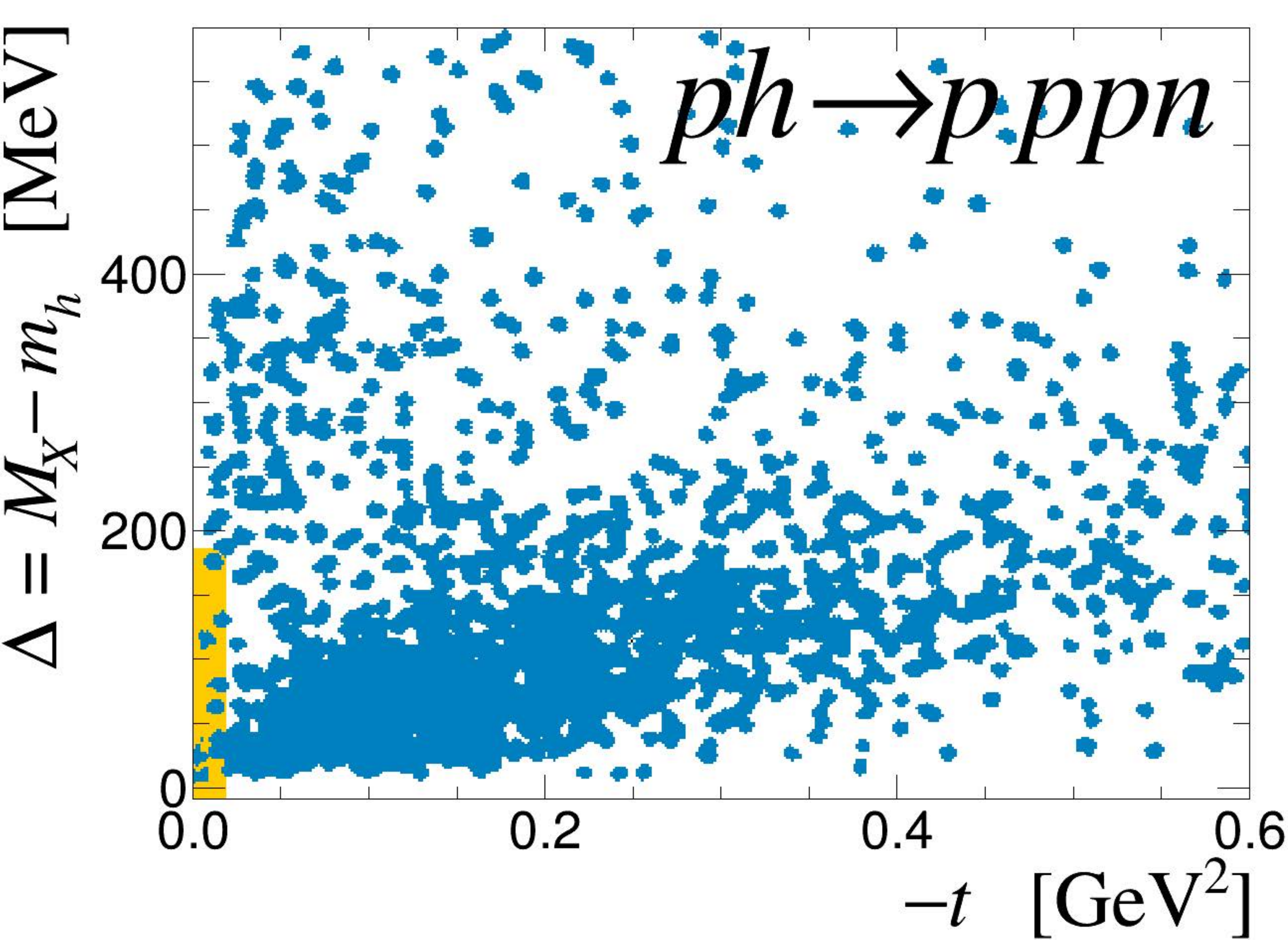}
  \includegraphics[width=0.9\columnwidth]{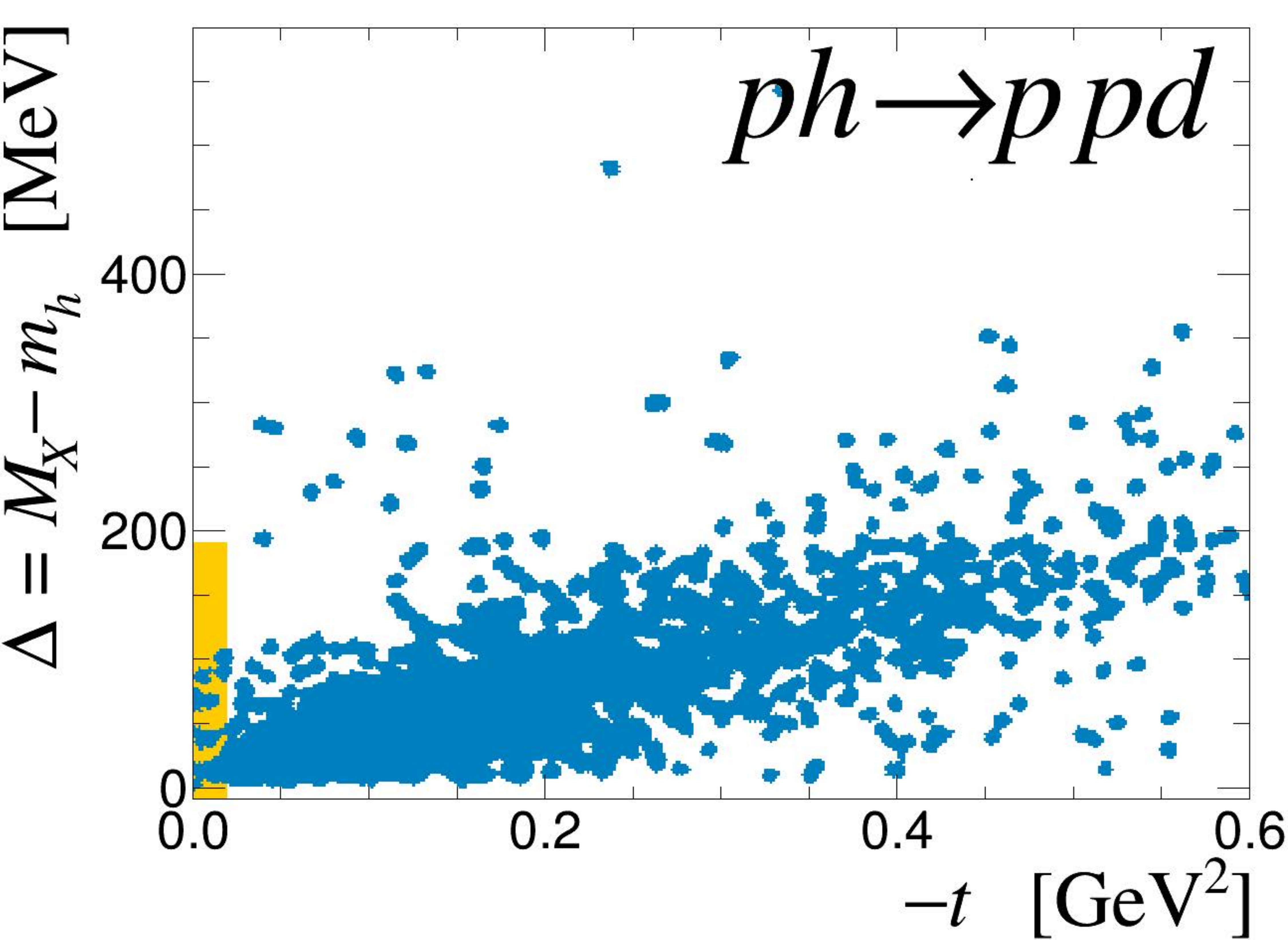}
  \end{center}
  \caption{\label{fig:ph}
    Scattered mass excess $\Delta$ versus momentum transfer plots for helion 4.6\,GeV/nucleon charge retention breakup scattering, $ph\!\to\!pppn$ and $ph\!\to\!ppd$, measured in the hydrogen bubble chamber experiment\,\cite{Stepaniak:1996sn}.  Filled areas indicate part of the distribution that can be studied at HJET. 
  }
\end{figure}

For a 4.6\,GeV/nucleon $hp$ scattering, the elastic, $\sigma_{h\to h}\!=\!24.2\!\pm\!1.0\,\text{mb}$, and breakup, $\sigma_{h\to pd}\!=\!7.29\!\pm\!0.14\,\text{mb}$ and $\sigma_{h\to ppn}\!=\!6.90\!\pm\!0.14\,\text{mb}$, cross sections were determined in the hydrogen bubble chamber measurements\,\cite{Dubna-Kosice-Moscow-Strasbourg-Tbilisi-Warsaw:1993lmp}. For the HJET momentum transfer range (\ref{eq:tRange}), the effective elastic cross section can be derived from the measured $\sigma_{h\to h}$:
\begin{equation}
  \sigma_{h\to h}^\text{HJET} \approx11\,\text{mb}.
\end{equation}

The scattered mass $M_X$ versus momentum transfer $t$ plots (see Fig.\,\ref{fig:ph}) for the breakup scattering were presented in Ref.\,\cite{Stepaniak:1996sn}. It may be pointed out that the correlation seen is in a {\em qualitative} agreement with Eq.\,(\ref{eq:Delta}). It was underlined in Ref.\,\cite{Stepaniak:1996sn} that the breakup events band is spread around the line
\begin{equation}
  M_X^2 = m_h^2 -2t,
\end{equation}
which is the same as that followed from Eq.\,(\ref{eq:Delta}) if $p_x\!=\!0$.

Each event in Fig.\,\ref{fig:ph} plots contributes nearly 0.003\,mb to the corresponding cross section. For $h\!\to\!ppn$, the breakup band events can be counted in the HJET momentum transfer range. After applying corrections due to the recoil proton detection efficiency at low $t$\,\cite{DUBNA-WARSAW:1975mxc}, one arrives to
\begin{equation}
  \sigma_{h\to ppn}^\text{HJET} < 0.02\,\text{mb}.
\end{equation}
Since, it is not unreasonable that mainly background events were counted, the result obtained should be interpreted as an upper limit.

Assuming that $h\!\to\!pd$ breakup has a flat $d\sigma/dt$ distribution in the $0<-t<0.45\,\text{GeV}^2$ momentum transfer range, one immediately finds $\sigma_{h\to pd}^\text{HJET}/\sigma_{h\to pd}\approx0.04$. However, since $\omega_h(t)\to0$ if $t\to0$, a correction is needed. Guessing that the correction is the same as for the $h\to pd$ distribution in Fig.\,\ref{fig:pd-ppn}, the effective cross-section can be estimated as
\begin{equation}
  \sigma_{h\to pd}^\text{HJET} \sim 0.15\,\text{mb}.
\end{equation}
For comparison, the result displayed in  Eq.\,(\ref{eq:hBreakup}) corresponds to about 0.25\,mb.

Thus, the following conclusion comes after Ref.\,\cite{Stepaniak:1996sn}:\\
\noindent{--~~} the helion beam breakup events which can be detected at HJET are mostly $h\to pd$;\\
\noindent{--~~} $\omega_h(T_R)$ depicted in Fig.\,\ref{fig:omega} should be interpreted as an upper limit for the helion beam breakup in HJET.

\begin{figure}[t]
  \begin{center}
  \includegraphics[width=0.73\columnwidth]{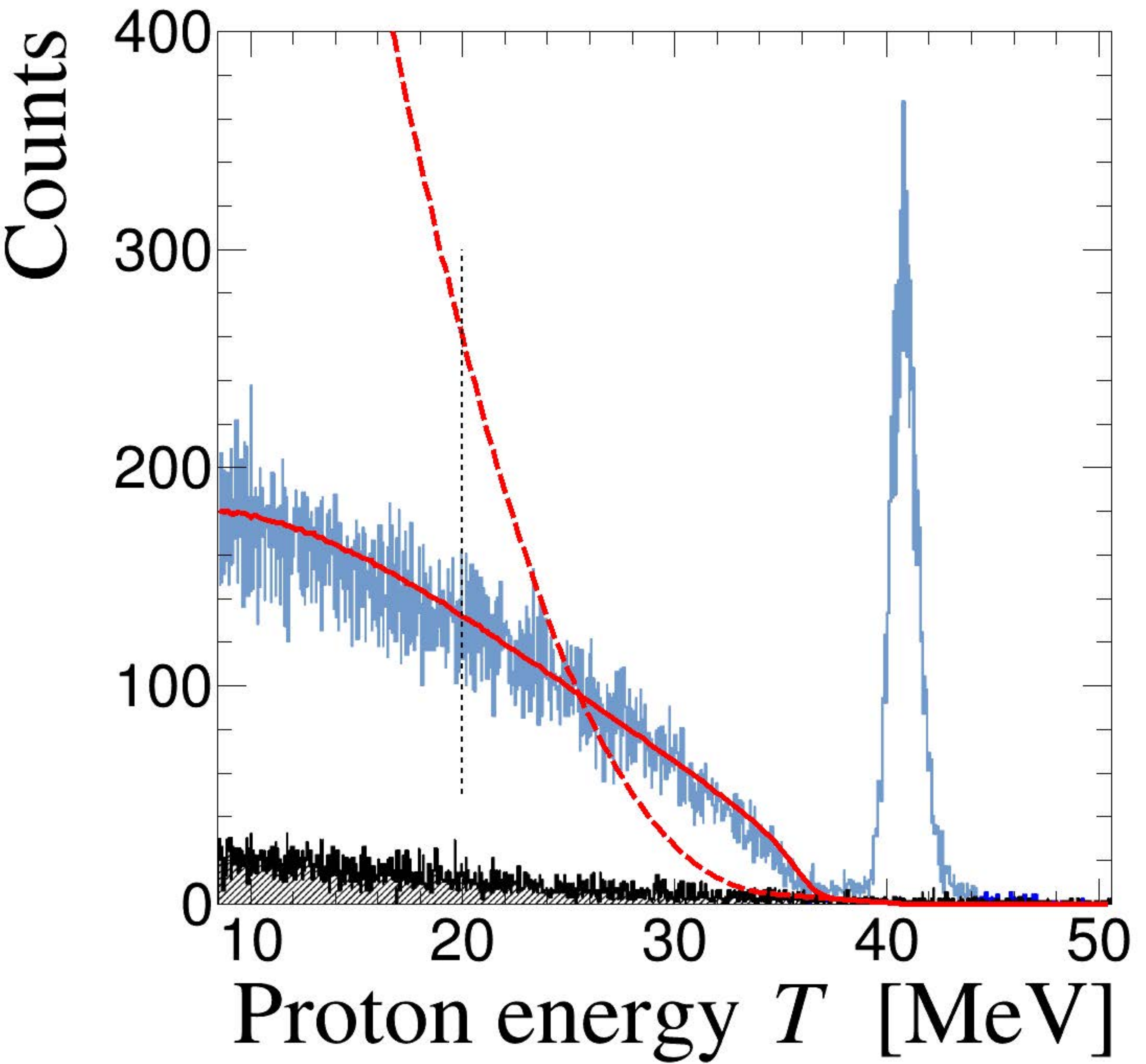}
  \end{center}
  \caption{\label{fig:65MeV}
    Light output spectrum of scattered protons obtained from the NaI(Tl) scintillator at $\theta_\text{lab}\!=\!70^\circ$ in the 65\,MeV $p\,{}^3\text{He}$ elastic scattering study\,\cite{Watanabe:2021jee}. The hatched region indicates the background contribution. Estimates (done here) for the $h\!\to\!{pd}$ and $h\!\to\!{ppn}$ breakup components are shown by solid and dashed red lines, respectively. The calculations were normalized by the total number of the breakup protons with energy greater than 20\,MeV ($\Delta\!<\!24\,\text{MeV}$).
  }
\end{figure}

Another evidence of the $h\!\to\!ppn$ breakup suppression for low $\Delta$ follows from the scattered proton energy spectrum (Fig.\,\ref{fig:65MeV}) measured in the 65\,MeV proton scattering off the $^3$He target\,\cite{Watanabe:2021jee}. The elastic events are centered at $T_\text{el}\!=\!40.8\,\text{MeV}$. Since energy measured in the breakup scattering can be roughly approached by $T_\text{brk}\!\approx\!T_\text{el}\!-\!\Delta$,  an evident square root dependence on $T$ at the breakup endpoint should be attributed to the phase space integral,  $\propto$\:\!$(\Delta\!-\!\Delta_\text{thr}^h)^{1/2}$, in the two-body breakup event rate in Eq.\,(\ref{eq:omegaTR-D}).

Applying Eqs.\,(\ref{eq:omegaTR-D})\,and\,(\ref{eq:omegaD0}) to the $h\!\to\!pd$ breakup and using $\sigma_p\!=\!90\,\text{MeV}$, one can fairly well approximate the experimental spectrum in Fig.\,\ref{fig:65MeV}. Since low energy, 65\,MeV, scattering is considered, the effective values of $\sigma_p$ and $|\bar{\psi}|$ may differ from those (\ref{eq:omegaCalib}) evaluated in the high energy, 10\:\!--\:\!31\,GeV, $pd$ scattering.

For the three-body breakup, $h\!\to\!ppn$, the phase space integral dependence in the event rate (\ref{eq:omegaTR-D}) is proportional to  $(\Delta\!-\!\Delta_\text{thr}^h)^2$, which makes the calculated spectrum inconsistent with the experimental one. So, there is no evidence of the $h\!\to\!ppn$ events in Fig.\,\ref{fig:65MeV}.

\section{Evaluation of the Breakup Corrections in the Helion Beam Polarization Measurements}

A basic assumption of the Glauber theory approach that a high energy beam proton crosses a target nucleus before any changes in the nucleus structure caused by the proton occur suggests that an electromagnetic interaction is the same for elastic and breakup scattering. However, to compare the electromagnetic amplitudes for the elastic and breakup scattering, the interaction, in accordance with Eq.\,(\ref{eq:Ffi}), should be projected to the final state wave functions $\Psi_h$ and $\Psi_{pd}$, respectively.
For $^3$He in the ground state, the function $f_\text{BW}(\Delta\!-\!\Delta_0,\sigma_\Delta)$ [Eq.\,(\ref{eq:omegaTR-D})], which is defined by the nucleon momentum distribution, is expected to be about the same for hadronic and electromagnetic scattering. Therefore, assuming that the disturbances of $^3$He , $^3\text{He}\to{^3\text{He}}^*(\bm{q})$, in the scatterings are small, it can be approximated that the reduced breakup to the elastic ratio (\ref{eq:omegaTR-D}) is
\begin{equation}
  \frac{\langle\bar{\Psi}_{pd}|{^3\text{He}}^*(\bm{q})\rangle_\text{em}}
       {\langle{\Psi}_{h} |{^3\text{He}}^*(\bm{q})\rangle_\text{em}}
       \approx
  \frac{\langle\bar{\Psi}_{pd}|{^3\text{He}}\rangle_\text{em}}
       {\langle{\Psi}_{h} |{^3\text{He}}\rangle_\text{em}}
       =
  \frac{\langle\bar{\Psi}_{pd}|{^3\text{He}}\rangle_\text{had}}
       {\langle{\Psi}_{h} |{^3\text{He}}\rangle_\text{had}}
       =
       \bar{\psi}.
       \label{eq:ratio}
\end{equation}
Consequently, the ratio for the electromagnetic amplitude (either spin-flip $\phi_5^\text{em}$  or non-flip $\phi_+^\text{em}$) must be nearly the same as for hadronic non-flip $\phi_+^\text{had}$, which leads to the following breakup corrections in the interference terms in Eq.\,(\ref{eq:AN}):
\begin{align}
  \phi_5^\text{em}\phi_+^\text{had}:\quad&\kappa \to \kappa +\kappa\times\omega_h(T_R), \label{eq:em5h+} \\
  \phi_5^\text{had}\phi_+^\text{em}:\quad&I_5 \to I_5 + \widetilde{I}_5\times\omega_h(T_R),\\
  \phi_5^\text{had}\phi_+^\text{had}:\quad&R_5 \to R_5 + \widetilde{R}_5\times\omega_h(T_R).
\end{align}

Consequently, correction (\ref{eq:xi}) to the measured beam polarization can be evaluated as
\begin{equation}
  \xi(T_R) =
  \frac{2\delta_{pd}}{\kappa_h}%
  \left(I_5 +R_5 T_R/T_c\right)\omega_h(T_R).
  \label{eq:xiTR}
\end{equation}

Assuming a linear fit of the measured polarization $P_\text{beam}(T_R)$ [see Eq.\,(\ref{eq:Pmeas})] in the  $2\!<\!T_R\!<\!10\,\text{GeV}$\,\cite{Poblaguev:2020Og} kinetic energy range, the breakup fraction functions can be replaced by
\begin{align}
  \omega_h(T_R) &\approx%
  \omega_0+\omega_1\frac{T_R}{T_c}=
  1.5\% + 0.3\%\,\frac{T_R}{T_c},
  \\
  \omega_h(T_R)\frac{T_R}{T_c} &\approx%
  \omega_0'+\omega_1'\frac{T_R}{T_c}=
  -9.6\% + 6.4\%\,\frac{T_R}{T_c}.
\end{align}

Thus, the breakup related systematic error in the beam polarization measurements can be evaluated as
\begin{equation}
  |\delta^\text{brk}P_h/P_h| =\left|
  \frac{2\delta_{pd}}{\kappa_h}\,(I_5\omega_0+R_5\omega_0')
  \right|
  <0.3\%,
  \label{eq:brkSyst}
  \end{equation}
in consistency with the EIC requirement (\ref{eq:systEIC}).

The relatively loose constraint on the systematic uncertainty can be explained by the essential non-linearity of the $\omega_h(T_R)T_R/T_c$ function. To improve, a parabolic function
\begin{equation}
  \xi(T_R) \approx \xi_0 + \xi_1\:\!T_R/T_c + \xi_2(T_R/T_c)^2
  \label{eq:brkPol2}
\end{equation}
can be considered to fit (assuming $\xi_0\!\equiv\!0$) $P_\text{beam}(T_R)$ in the  $2\!<\!T_R\!<\!7\,\text{GeV}$ energy range. In this case
\begin{equation}
  \omega_0=0.28\%,\qquad\omega_0'=-0.21\%
\end{equation}
and the breakup related uncertainty (\ref{eq:brkSyst}) is diluted to a negligible value.

To provide better control of the possible breakup corrections, one can consider the following function 
\begin{equation}
  \xi(T_R) \approx \xi_1{T_R}/{T_c}%
  + \left[\xi_0^\omega + \xi_1^\omega{T_R}/{T_c}\right]\times\omega_h(T_R)
  \label{eq:brkFit}
\end{equation}
to fit the measured polarization. A more accurate estimate of $\omega_h(T_R)$ can be done, e.g., in (relatively) low energy, 5\:\!--\:\!20\,GeV, $^3$He beam measurements at HJET or by tagging the $^3$He breakup by detecting the spectator nucleons in dedicated downstream detectors\,\cite{Nunes:2022cmc}.

It should be noted that the breakup corrections to the dominant interference term  $\phi_5^\text{em}\phi_+^\text{had}$ (\ref{eq:em5h+}) are relatively large, up to 4\%. However, the effect cancels in the analyzing power ratio. Under a more general assumption, the correction may be not the same as in the non-flip scattering, $\omega_h(T_R)\!\times\!(1+c)$, and may be different for the $p^\uparrow{h}$ and $h^\uparrow{p}$ scattering. In this case, the corresponding systematic error in the beam polarization measurement is $(c_{ph}\!-\!c_{hp})\omega_0$ and may be considered as a small one, $\ll$\;\!1\%, even if $|c_{ph}\!-\!c_{hp}|\!\sim\!1$. 

Eq.\,(\ref{eq:em5h+}) was derived assuming $t\!\to\!0$. In the dimension based analysis, the upper limit for the corresponding systematic error in polarization measurement can be evaluated as $|B^{ph}t_c^{ph}\omega_0'|\!\ll\!0.1\%$. Here,  $B^{ph}\!=\!33\,\text{GeV}^{-2}$\,\cite{Dubna-Kosice-Moscow-Strasbourg-Tbilisi-Warsaw:1993lmp} is the elastic $ph$ slope parameter.

Applying Eq.\,(\ref{eq:ratio}) to the $p^\uparrow{h}$ spin-flip amplitude, one readily finds $\widetilde{r}_5^{ph}\!=\!r_5^{ph}$ in agreement with Eqs.\,(\ref{eq:r5el}) and (\ref{eq:r5brk}). Similarly, for the $h^\uparrow{p}$ spin-flip amplitude $\widetilde{r}_5^{hp}\!=\!r_5^{hp}$, i.e., $\delta_{pd}\!=\!0$. If so, see Eq.\,(\ref{eq:brkSyst}), the breakup corrections completely cancel in the $^3$He beam polarization measurements with HJET.

\section{Summary}

In this paper, evaluation\,\cite{Poblaguev:2022hsi} of the breakup corrections in the EIC $^3$He beam polarization measurements with HJET was reviewed.

For proton-deuteron scattering, the breakup fraction evaluated at HJET\,\cite{Poblaguev:2022hsi} was found to be consistent with a value obtained in the hydrogen bubble chamber experiment\,\cite{Aladashvili:1977xe}. This result proves, that within the experimental accuracy of the measurements\,\cite{Poblaguev:2022hsi}, \emph{(i)} the model used to describe the breakup event rate, $dN/dT_Rd\Delta$, adequately simulates the experimental data and \emph{(ii)} the breakup event fraction can be reliably monitored in the HJET measurements using 5\:\!--\:\!20\,GeV $^3$He beams.

In Ref.\,\cite{Poblaguev:2022hsi}, the breakup fraction in the future $^3$He beam polarization measurement at EIC was estimated by extrapolating the $d\!\to\!pn$ results to the two-body $h\!\to\!pd$ breakup and neglecting the three-body $h\!\to\!ppn$ one. Analyzing experimental distributions displayed in Refs.\,\cite{Stepaniak:1996sn,Watanabe:2021jee}, it was confirmed that the $h\!\to\!ppn$ rate is very small in the HJET measurements.

The evaluated\,\cite{Poblaguev:2022hsi} $^3$He breakup rate  was compared with the results of the $^3$He beam scattering in the hydrogen bubble chamber\,\cite{Dubna-Kosice-Moscow-Strasbourg-Tbilisi-Warsaw:1993lmp,Stepaniak:1996sn}. It was found that $\omega_h(T_R)$ displayed in Fig.\,\ref{fig:omega} should be interpreted as an upper limit for the breakup fraction in the EIC $^3$He beam scattering in HJET.

To estimate breakup effects in the $^3$He beam polarization measurements, corrections to the interference terms in Eq.\,(\ref{eq:AN}) are needed to be known for both $p^\uparrow h$ and $h^\uparrow p$ scattering.

It was shown that, within the applicability of the Glauber theory, the ratio of the proton-nucleus breakup spin-flip and non-flip amplitudes is the same as for elastic proton-proton scattering.

Regardless the breakup corrections to the $\phi_5^\text{em}\phi_+^\text{had}$ and $\phi_5^\text{had}\phi_+^\text{em}$ terms in Eq.\,(\ref{eq:AN}) were underestimated in Ref.\,\cite{Poblaguev:2022hsi}, this did not lead to a wrong conclusion, as the miscalculation of up to an order of magnitude was being tolerated in the analysis done.

Although the corrections to the interference terms, found here, are relatively large, up to 4\%, they cancel to a negligible level in the analyzing power ratio in Eq.\,(\ref{eq:Ph}).

To recognize a residual breakup correction (if any) in experimental data analysis, the measured polarization $P_\text{meas}(T_R)$ (\ref{eq:Pmeas}) should be interpolated using $\xi(T_R)$ given by Eq.\,(\ref{eq:brkFit}) or, in simplified consideration, by Eq.\,(\ref{eq:brkPol2}). 
\acknowledgements{
  The author would like to thank B. Z. Kopeliovich for useful discussions and acknowledges support from the Office of Nuclear Physics in the Office of Science of the US Department of Energy. This work is authored by employees of Brookhaven Science Associates, LLC under Contract No.\,DE-SC0012704 with the U.S. Department of Energy.}

\end{document}